\documentclass[a4paper,12pt]{article}
\usepackage{amsfonts}
\usepackage{latexsym}
\usepackage{amssymb}
\date{\empty}

\begin{document}

\title{Second Law and
Non-Equilibrium Entropy\\ of Schottky Systems\\
--Doubts and Verification--}
\author{W. Muschik\footnote{Corresponding author:
muschik@physik.tu-berlin.de}
\\
Institut f\"ur Theoretische Physik\\
Technische Universit\"at Berlin\\
Hardenbergstr. 36\\D-10623 BERLIN,  Germany}
\maketitle

            \newcommand{\be}{\begin{equation}}
            \newcommand{\beg}[1]{\begin{equation}\label{#1}}
            \newcommand{\ee}{\end{equation}\normalsize}
            \newcommand{\bee}[1]{\begin{equation}\label{#1}}
            \newcommand{\bey}{\begin{eqnarray}}
            \newcommand{\byy}[1]{\begin{eqnarray}\label{#1}}
            \newcommand{\eey}{\end{eqnarray}\normalsize}
            \newcommand{\beo}{\begin{eqnarray}\normalsize}
         
            \newcommand{\R}[1]{(\ref{#1})}
            \newcommand{\C}[1]{\cite{#1}}

            \newcommand{\mvec}[1]{\mbox{\boldmath{$#1$}}}
            \newcommand{\x}{(\!\mvec{x}, t)}
            \newcommand{\m}{\mvec{m}}
            \newcommand{\F}{{\cal F}}
            \newcommand{\n}{\mvec{n}}
            \newcommand{\argm}{(\m ,\mvec{x}, t)}
            \newcommand{\argn}{(\n ,\mvec{x}, t)}
            \newcommand{\T}[1]{\widetilde{#1}}
            \newcommand{\U}[1]{\underline{#1}}
            \newcommand{\V}[1]{\overline{#1}}
            \newcommand{\ub}[1]{\underbrace{#1}}
            \newcommand{\X}{\!\mvec{X} (\cdot)}
            \newcommand{\cd}{(\cdot)}
            \newcommand{\Q}{\mbox{\bf Q}}
            \newcommand{\p}{\partial_t}
            \newcommand{\z}{\!\mvec{z}}
            \newcommand{\bu}{\!\mvec{u}}
            \newcommand{\rr}{\!\mvec{r}}
            \newcommand{\w}{\!\mvec{w}}
            \newcommand{\g}{\!\mvec{g}}
            \newcommand{\D}{I\!\!D}
            \newcommand{\se}[1]{_{\mvec{;}#1}}
            \newcommand{\sek}[1]{_{\mvec{;}#1]}}            
            \newcommand{\seb}[1]{_{\mvec{;}#1)}}            
            \newcommand{\ko}[1]{_{\mvec{,}#1}}
            \newcommand{\ab}[1]{_{\mvec{|}#1}}
            \newcommand{\abb}[1]{_{\mvec{||}#1}}
            \newcommand{\td}{{^{\bullet}}}
            \newcommand{\eq}{{_{eq}}}
            \newcommand{\eqo}{{^{eq}}}
            \newcommand{\f}{\varphi}
            \newcommand{\rh}{\varrho}
            \newcommand{\dm}{\diamond\!}
            \newcommand{\seq}{\stackrel{_\bullet}{=}}
            \newcommand{\st}[2]{\stackrel{_#1}{#2}}
            \newcommand{\om}{\Omega}
            \newcommand{\emp}{\emptyset}
            \newcommand{\bt}{\bowtie}
            \newcommand{\btu}{\boxdot}
            \newcommand{\tup}{_\triangle}
            \newcommand{\tdo}{_\triangledown}
            \newcommand{\Ka}{\frac{\nu^A}{\Theta^A}}
            \newcommand{\K}[1]{\frac{1}{\Theta^{#1}}}
            \newcommand{\ap}{\approx}
            \newcommand{\bg}{\st{\Box}{=}}
            \newcommand{\si}{\simeq}
\newcommand{\Section}[1]{\section{\mbox{}\hspace{-.6cm}.\hspace{.4cm}#1}}
\newcommand{\Subsection}[1]{\subsection{\mbox{}\hspace{-.6cm}.\hspace{.4cm}
\em #1}}

\newcommand{\const}{\textit{const.}}
\newcommand{\vect}[1]{\underline{\ensuremath{#1}}}  
\newcommand{\abl}[2]{\ensuremath{\frac{\partial #1}{\partial #2}}}

\vspace{.5cm}

\noindent
{\bf Abstract} 
Meixner's historical remark in 1969 "... it can be shown that the concept of entropy in the absence of equilibrium is in fact not only questionable but that it cannot even be defined...."
is investigated from today's insight. Several statements --such as the three laws of
phenomenological thermodynamics, the embedding theorem and the adiabatical uniqueness--
are used to get rid of non-equilibrium entropy as a primitive concept. In this framework,
Clausius inequality of open systems can be derived by use of the defining inequalities which establish the non-equilibrium quantities contact temperature and non-equilibrium molar entropy which allow to describe the interaction between the Schottky system and its controlling
equilibrium environment.

\section{Introduction}

The Second Law has many faces: there are different formulations in phenomenological thermodynamics, statistics, kinetics and quantum theory. Here, only phenomenological
considerations are made in the range of Schottky systems \C{SCHO29} --that are discrete
systems which can exchange heat, power and material with their environment by suitable
partitions--. The field formulation of thermodynamics is as well as a historical survey and an axiomatic treatment out of scope.   
\vspace{.3cm}\newline
Historically, the Second Law launches with two verbal formulations concerning irreversible cyclic
processes of discrete systems: the principle of Kelvin \C{Kelvin} and that of Clausius \C{Clausius}
which are so well known that they can be formulated in a short form \C{MUAS}:
\begin{center}
\parbox[t]{14cm}{Kelvin:\quad There is no Thomson process (but friction processes exist),\newline
Clausius: There is no Clausius process (but heat conduction processes exit).}
\end{center}
Accepting additionally
\begin{center}
\parbox[t]{14cm}{Carnot \C{Carnot}: Reversible Carnot processes exist\\ $\mbox{}$\hspace{2cm}
(not really, but as a mathematical closure of irreversible processes),}
\end{center}
and starting with these verbal statements of Kelvin, Clausius and Carnot,
the following Clausius inequality valid for cyclic processes in closed systems can be derived \C{MU89}
in an up-to-date formulation 
\bee{1}
\oint\frac{\st{\td}{Q}(t)}{T^\Box(t)}dt\ \leq\ 0.
\ee
Here, $\st{\td}{Q}$ is the heat exchange per time between the controlling heat reservoir and
the Schottky system which can be measured by calorimetry. $T^\Box$
is the thermostatic (equilibrium) temperature of the heat reservoir which controls the cyclic process. 
\vspace{.3cm}\newline
For more detailed understanding of Clausius inequality, the following questions arise and have
to be discussed below
\begin{enumerate}
\item What is the meaning of the parameter $t$ in connection with the $<$ and $=$ signs in
Clausius inequality ?\vspace{-.3cm}
\item On what state space characterizing the system operates the cyclic process ?\vspace{-.3cm}
\item How to extend the inequality to open systems ?\vspace{-.3cm}
\item What is the relationship between Clausius inequality and entropies ?
\end{enumerate}
Beyond these questions, a shortcoming of the derivation of Clausius inequality \R{1} has to be taken
into consideration: the statement Carnot claims the existence of reversible processes, a
presupposition which should not be used here, because the physical meaning of reversible processes 
is not evident and has to be defined properly in the course of this paper. Additionally, the Carnot
theorem of reversible Carnot processes 
\bee{2}
\frac{\st{\td}{Q}_1\Delta_1t}{T^\Box_1}+\frac{\st{\td}{Q}_2\Delta_2t}{T^\Box_2}\ =\
\frac{Q_1}{T^\Box_1}+\frac{Q_2}{T^\Box_2}\ =\ 0
\ee
which is used in the derivation of \R{1} is a special case of this relation which should be proved.
Consequently, the verbal formulation of the Second Law --statements Kelvin and Clausius--
cannot be transformed by the statement Carnot into Clausius inequality without a logical fallacy.
A remedy may be to set \R{1} as an axiom, or better, to derive it in connexion with a suitable
definition of a non-equilibrium entropy, a way which is worked out here serving as a motivation
to have a look at this "antiquated stuff" again.
\vspace{.3cm}\newline
Starting point is that the sign of the LHS of \R{1} is unknown. Using this fact, question \#3 can be
taken into attac by investigating the following expression
\bee{3} 
SL\ :=\ \oint\Big(\frac{\st{\td}{Q}(t)}{T^\Box(t)}
+ \mvec{s}^\Box\cdot\st{\td}{\mvec{n}}\!{^e}\Big)dt.
\ee
Here, $\st{\td}{\mvec{n}}\!{^e}$ is the mole numbers exchange between the controlling reservoir
and the system, and $\mvec{s}^\Box$ is the molar equilibrium entropy of this reservoir. $SL$ has a characteristic shape: $\st{\td}{Q}$ and $\st{\td}{\mvec{n}}\!{^e}$ are exchange quantities
referring to the system, whereas $T^\Box$ and $\mvec{s}^\Box$ belong to the controlling reservoir
which generates the cyclic process of the system.
\vspace{.3cm}\newline
Although still in use today, Clausius inequality is a historical relation.
In the meantime, there are "as many formulations of the Second Law as there are authors"
\C{Hutter}. Some of these formulations can be found in
\C{KE76,SER79,SIL83,MU88,MU90b,MUEH96,MU04}.

\section{Schottky Systems}

\subsection{Exchanges and partitions}

A system $\cal G$, described as undecomposed and homogeneous,
which is separated by a partition $\partial {\cal G}$ from its environment
${\cal G}^\Box$ is called a {\em Schottky system} \C{SCHO29}, if the interaction
between $\cal G$ and ${\cal G}^\Box$ through $\partial {\cal G}$ can be described by
\bee{4}
\mbox{heat exchange\ }\st{\td}{Q},\quad\mbox{power exchange\ }\st{\td}{W},\quad
\mbox{ and material exchange\ }\st{\td}{\mvec{n}}\!{^e}.
\ee
The power exchange is related to the work variables $\mvec{a}$ of the system
\bee{5}
\st{\td}{W}\ =\ {\bf A}\cdot\st{\td}{\mvec{a}}.
\ee
Here, ${\bf A}$ are the generalized forces which are as well known as the work variables.
Kinetic and potential energy are constant and therefore out of scope.
$\st{\td}{Q}$ is measurable by calorimetry and the time rate of the mole numbers due to
material exchange $\st{\td}{\mvec{n}}\!{^e}$ by weigh.\nolinebreak
\vspace{.3 cm}\newline
Using the exchange quantities, partitions $\partial {\cal G}$ of different properties can be defined:
If by choice of an arbitrary environment ${\cal G}^\Box$, the following exchange quantities vanish identically 
\byy{6}
\st{\td}{W}\ \equiv\ 0&\longrightarrow&\mbox{power-isolating},
\\ \label{7}
\st{\td}{\mvec{n}}\!{^e}\ \equiv\ \mvec{0}&\longrightarrow&\mbox{material-isolating},
\\ \label{8}
\st{\td}{Q}\ \equiv\ 0\ \wedge\ 
\st{\td}{\mvec{n}}\!{^e}\ \equiv\ \mvec{0}&\longrightarrow&\mbox{adiabatic},
\\ \label{9}
\mbox{adiabatic and power-isolating}&\longrightarrow&\mbox{isolating},
\eey
the partition is called $\longrightarrow\boxtimes$.
A system is called {\em thermally homogeneous}, if it does not contain any adiabatic
partition in its interior.
\vspace{.3 cm}\newline
An {\em inert partition} does not absorb or emit heat, power and material \C{MU09}.
It is defined by the following equations \C{MUBER04,MUBER07}
\bee{9a}  
\st{\td}{Q}\ =\ -\st{\td}{Q}\!{^\Box},\qquad
{\bf A}\cdot\st{\td}{\mvec{a}}\ =\ {\bf A}^\Box\cdot\st{\td}{\mvec{a}},\qquad
\st{\td}{\mvec{n}}\!{^e}\ =\ -\st{\td}{\mvec{n}}\!{^{e\Box}}.
\ee
Here, the $^\Box$-quantities belong to the system's controlling environment ${\cal G}^\Box$.
The work done on the system is
performed by the environment using its generalised forces ${\bf A}^\Box$ and
orientated at the work variables of the system. The permeability of $\partial\cal G$ to 
power and material is described by \R{9a}$_{2,3}$.

\subsection{State spaces and processes}

The cyclic process $SL$ in \R{3} is defined on a state space because otherwise the path integral
makes no sense. Here, a large state space $\cal Z$ is used
\C{MUAS1} which is decomposed into its equilibrium subspace and the non-equilibrium part
\bee{10}
Z\ =\ ({\sf z}_{eq},{\sf z}_{ne})\ \in\ {\cal Z}.
\ee
Here, {\em states of equilibrium} ${\sf z}_{eq}$ are defined by time independent states of an isolated Schott\-ky system. The equilibrium subspace is too small for decribing
non-equilibrium. Consequently, it has to be extended by the non-equilibrium part
${\sf z}_{ne}$ of ${\cal Z}$. If the considered system is in equilibrium, the
non-equilibrium variables become dependent on the equilibrium ones
\bee{10a}
Z^{eq}\ =\ \Big({\sf z}_{eq},{\sf z}_{ne}({\sf z}_{eq})\Big)\ \in\ {\cal Z}^{eq}.
\vspace{.3 cm}\ee
The variables of the equilibrium subspace are determined by the {\em Zeroth Law}:
The state space of a thermal homogeneous Schottky system in equiIibrium is spanned by the work variables, the mole numbers and the internal energy
\bee{11}
{\sf z}_{eq}\ =\ (\mvec{a},\mvec{n}, U)\quad\longrightarrow\quad
Z\ =\ (\mvec{a},\mvec{n}, U,{\sf z}_{ne}).
\ee
A projection $\cal P$ is introduced which projects the
non-equilibrium state $Z$ onto the equilibrium subspace
\bee{19w}
{\cal P}Z\ =\ {\cal P}\Big(\mvec{a},\mvec{n}, U,{\sf z}_{ne}\Big)\ =\
\Big(\mvec{a},\mvec{n}, U\Big)\ :=\ Z^*
\ee
whose equilibrium states are marked by $^*$. 
A process
\bee{20w}
Z(t)\ =\ \Big(\mvec{a},\mvec{n}, U,{\sf z}_{ne}\Big)(t),\quad t=\mbox{time}
\ee
generates by projection a trajectory on the equilibrium subspace
\bee{21w}
{\cal P}Z(t)\ \equiv\ Z^*(t)\ =\ \Big(\mvec{a},\mvec{n}, U\Big)(t),\qquad {\sf z}_{ne}(\mvec{a},\mvec{n}, U)
\ee
which is called a {\em reversible process},
a bit strange denotation because no "process" with progress in time takes place on the
equilibrium subspace. The "time" in
\R{21w} is generated by projection and represents the path parameter along the reversible
process. $Z^*(t)$ is also denoted as the {\em accompanying process of Z(t)} \C{KE71}.
Although not existing in nature, reversible processes serve as mathematical closing of the
"real" (irreversible) processes which are defined as trajectories on the non-equilibrium state space.

\subsection{The First Law}

Up to now, the internal energy was introduced in \R{11}$_1$ as one variable of the quilibrium
subspace of a thermally homogeneous Schottky system. The connection between the time rate
of the internal energy of the system and the exchange quantities through $\partial\cal G$ is
establiched by the {\em First Law}
\bee{17}
\st{\td}{U}\ =\ \st{\td}{Q} + \mvec{h}\cdot\st{\td}{\mvec{n}}\!{^e} + \st{\td}{W} 
\ee
which states that the internal energy $U$ of the system should be conserved in isolated
Schottky systems. The second term of the RHS of \R{17} originates from the fact that the heat
exchange has to be redefined for open systems ($\st{\td}{\mvec{n}}\!{^e}\neq\mvec{0}$) 
\C{MUGU99}. Here, $\mvec{h}$ are the molar enthalpies of the chemical components in $\cal G$.
The modified heat exchange which is combined with the material exchange
appearing in the First Law \R{17} was used by R. Haase \C{HA69}.
\vspace{.3cm}\newline
Because $U$ is not defined as a state function, but rather as a state
variable, $\st{\td}{U}$ is not defined as a total differential in \R{17}.
Internal mole number changes
\bee{18}
\st{\td}{\mvec{n}}\!{^i}\ =\ \st{\td}{\mvec{n}}-\st{\td}{\mvec{n}}\!{^e}
\ee
by chemical reactions are not influencing the internal energy which
is also conserved in isolated systems undergoing chemical reactions \C{HA69}.
How to define the internal energy in more detail can be found in \C{BO21}.  
\vspace{.3cm}\newline
In equilibrium systems, the internal energy is connected with the thermostatic equilibrium
temperature by the caloric equation of state. Such a relation is missing in non-equilibrium sytems, because a non-equilibrium temperature is not unequivocally defined. But fact is, that also in
non-equilibria temperatures are measured, and the question arises, what is the nature of these temperatures (do not think of perfect gases).

\section{The Second Law}

\subsection{Doubts: Non-equilibrium entropy}

Once upon a time, Meixner wrote in 1967 \C{MEI67}:..."The idea that an unambiguous entropy also exists
in the absence of equilibrium, likewise propounded by Clausius, has been accepted almost entirely
without further examination and applied with no little success. An analysis of Clausius' work
however reveals an inexactitude in the logic of his deductions which cannot be overlooked.
With the aid of thermodynamic systems of the simplest kind, namely electrical networks, it can
be shown that the concept of entropy in the absence of equilibrium is in fact not only questionable
but that it cannot even be defined. This leads to the problem of developing a thermodynamic
theory of processes....which is disassociated from the concept of entropy in the absence of
equilibrium. This latter can be achieved by applying the principle of the fundamental inequality
which represents an interpretation of the Second Law of non-equilibrium thermodynamics,
disassociated from the concept of entropy."
\vspace{.3cm}\newline
More details concerning the "entropy-free" non-equilibrium thermodynamics which replaces the
"dubious non-equilibrium entropy" by the Fundamental Inequality can be found in
\C{MEI68, MEI69,KEL72,KER72}. 
\vspace{.3cm}\newline
Despite of Meixner's warning, the thermodynamic society, especially that of cele\-bra\-ting
Rational Thermodynamics uses successfully non-equilibrium entropies and also "non-equilibrium
temperature" as primitive concepts. Therefore the question arises, whether there is a possibility
to create a scheme for constructing non-equilibrium entropies unequivocally? Evident is, that the
concept of thermostatic equilibrium temperature has to be adapted to non-equilbrium because
the differential of a non-equilibrium entropy contains temperature.

\subsection{Defining inequalities}

Consider a discrete system $\cal G$ and its environment ${\cal G}^\Box $ which are separated from each
other by a partition $\partial\cal G$.
\begin{center}
\parbox[t]{14cm}{{\sf Definition:} A quantity $\bf J$ of ${\cal G}$ is called {\em balanceable}, if its time rate
can be decomposed into a flux $\bf\Psi$ through $\partial\cal G$ and a production $\bf R$ in
$\cal G$
\bee{19}
\st{\td}{\bf J}\ =\ {\bf\Psi} + {\bf R},\qquad {\bf\Psi}\ =\
{\bf\Phi}+ \varphi\st{\td}{\mvec n}\!{^e}.
\vspace{-.3cm}\ee}
\end{center}
The flux is composed of its conductive part $\bf\Phi$ and its convective part
$\varphi\st{\td}{\mvec n}\!{^e}$. Setting
\begin{center}
\parbox[t]{14cm}{
{\sf Axiom I:} The entropy is in equilibrium and in non-equilibrium a balanceable quantity.}
\end{center}
The equilibrium environment ${\cal G}^\Box $ is presupposed to be a
reservoir, that means, that the relaxation times are arbitrary high and that ${\cal G}^\Box $
can be described as being always in equilibrium. Consequently, ${\cal G}^\Box $ is subjected to
thermostatics whose validity is presupposed. According to axiom I, the time rate of the entropy of
the controlling reservoir is
\bee{20}
\st{\td}{S}\!{^\Box}\ =\ \frac{1}{T^\Box}\st{\td}{Q}\!{^\Box}+
\mvec{s}{^\Box}\cdot\st{\td}{\mvec n}\!^{\Box e}.
\ee
Here, the entropy flux is a factorized decomposition into the reciprocal thermostatic
temperature $T^\Box$ of the environment and the heat exchange through $\partial\cal G$.
Also the components of the external material exchange $\st{\td}{\mvec n}\!^{\Box e}$ are in
reference to the environment. The molar entropies of the components in ${\cal G}^\Box$
are $\mvec{s}{^\Box}$. An entropy production does not appear in \R{20}, because
${\cal G}^\Box $ is an equilibrium system and consequently, all thermostatic quantities are defined.
\vspace{.3cm}\newline
According to axiom I, a non-equilibrium entropy --which has to be defined in the sequel-- has the
form
\bee{21}
\st{\td}{S}\ =\ \frac{1}{\Theta}\st{\td}{Q}+\mvec{s}\cdot\st{\td}{\mvec n}\!{^e}+\Sigma.
\ee
Up to the exchange quantities through $\partial\cal G$, $\st{\td}{Q}$ and
$\st{\td}{\mvec n}\!{^e}$ --which are fixed due to the inert partition according to
\R{9a}$_{1,3}$-- the temperature $\Theta$,
the molar entropies $\mvec{s}$ and the entropy production $\Sigma$ are unknown and have to be defined in the sequel. The balancebility \R{21} of a non-equilibrium entropy represents the first
step for defining expressions which are beyond Meixner's criticism. But first of all, 
$\Theta$, $\mvec{s}$ and $\Sigma$ are only place holders in the calculation.
Setting
\begin{center}
\parbox[t]{14cm}{
{\sf Axiom II:} The entropies of partial systems are additive.}
\end{center}
The entropy of the isolated total system ${\cal G}^\Box \cup {\cal G}$ is according to axoim II
\bey\nonumber
\st{\td}{S}\!^{tot}\ =\ \st{\td}{S} + \st{\td}{S}\!{^\Box} &=&
\frac{1}{\Theta}\st{\td}{Q}+\mvec{s}\cdot\st{\td}{\mvec n}\!{^e}+
\frac{1}{T^\Box}\st{\td}{Q}\!{^\Box}+
\mvec{s}{^\Box}\cdot\st{\td}{\mvec n}\!^{\Box e}+\Sigma\ =\
\\ \label{25}
&=&\Big(\frac{1}{\Theta}-\frac{1}{T^\Box}\Big)\st{\td}{Q}
+(\mvec{s}-\mvec{s}{^\Box})\cdot\st{\td}{\mvec n}\!{^e}+\Sigma,
\eey
by inserting the properties \R{9a}$_{1,3}$ of the inert $\partial\cal G$.
\vspace{.3cm}\newline
Setting
\begin{center}
\parbox[t]{14cm}{
{\sf Axiom III:} The Second Law for isolated systems --here ${\cal G}^\Box \cup {\cal G}$--
\bee{26}
\st{\td}{S}\!^{tot}\ \geq\ 0
\ee}
\end{center}
results according to \R{25} in
\bee{27}
\Big(\frac{1}{\Theta}-\frac{1}{T^\Box}\Big)\st{\td}{Q}
+(\mvec{s}-\mvec{s}{^\Box})\cdot\st{\td}{\mvec n}\!{^e}+\Sigma\ \geq\ 0.
\ee
Presupposing that all chemical components have the same temperature --$\Theta$ in $\cal G$ and
$T^\Box$ in ${\cal G}^\Box$-- the molar entropies $\mvec{s}$ can be decomposed into molar
enthalpies $\mvec{h}$ and chemical potentials $\mvec{\mu}$ \C{KES79}
\bee{31z}
\mvec{s}{^\Box}\ =\ \frac{1}{T^\Box}\Big(\mvec{h}^\Box-\mvec{\mu}^\Box\Big),\qquad
\mvec{s}\ =\ \frac{1}{\Theta}\Big(\mvec{h}-\mvec{\mu}\Big),
\ee
and \R{27} results in the {\em dissipation inequality}
\bee{27a}
\Big(\frac{1}{\Theta}-\frac{1}{T^\Box}\Big)\st{\td}{Q}
+\Big(\frac{\mvec{h}}{\Theta}-\frac{\mvec{h}^\Box}{T^\Box}\Big)\cdot\st{\td}{\mvec n}\!{^e}
+\Big(\frac{\mvec{\mu}^\Box}{T^\Box}-\frac{\mvec{\mu}}{\Theta}\Big)\cdot\st{\td}{\mvec n}\!{^e}
+\Sigma\ \geq\ 0.
\ee
Setting
\begin{center}
\parbox[t]{14cm}{
{\sf Axiom IV:} The Second Law for entropy productions --here in $\cal G$--
\bee{28}
\Sigma\ \geq\ 0.
\ee}
\end{center}
Up to now, $\Theta$, $\mvec{h}$ and $\mvec{\mu}$ are place holders in the dissipation inequality
\R{27a}. By setting
\begin{center}
\parbox[t]{14cm}{
{\sf Axiom V:} In accordance with the dissiparion inequality \R{27a}, 
{\em defining inequa\-li\-ties} are demanded for introducing the place holders
$\Theta$, $\mvec{h}$ and $\mvec{\mu}$ of $\cal G$ 
\bee{29}
\Big(\frac{1}{\Theta}-\frac{1}{T^\Box}\Big)\st{\td}{Q}\ \st{*}{\geq}\ 0,\quad
\Big(\frac{\mvec{h}}{\Theta}-\frac{\mvec{h}^\Box}{T^\Box}\Big)\cdot\st{\td}{\mvec n}\!{^e}\ 
\st{*}{\geq}\ 0,\quad
\Big(\frac{\mvec{\mu}^\Box}{T^\Box}-\frac{\mvec{\mu}}{\Theta}\Big)\cdot\st{\td}{\mvec n}\!{^e}\
\st{*}{\geq}\ 0,
\ee}
\end{center}
as discussed in the next section.

\section{Contact Quantities}

First of all, the following proposition \C{MU84} is used:
\bee{30}
{\bf X}\cdot f({\bf X})\ \leq\ {\bf 0}\ (\mbox{for all}\ {\bf X}\wedge f \ \mbox{continuous at}\ 
{\bf X} = {\bf 0})\ \Longrightarrow\ f({\bf 0}) = {\bf 0}
\ee
which is applied to \R{29} for an inert partition \R{9a}$_{1,3}$ 
\bee{31}
\Big(\frac{1}{\Theta}-\frac{1}{T^\Box}\Big)\st{\td}{Q}\!{^\Box}\ \leq\ 0,\quad
\Big(\frac{\mvec{h}}{\Theta}-\frac{\mvec{h}^\Box}{T^\Box}\Big)\cdot\st{\td}{\mvec n}\!{^{\Box e}}\ 
\leq\ 0,\quad
\Big(\frac{\mvec{\mu}^\Box}{T^\Box}-\frac{\mvec{\mu}}{\Theta}\Big)\cdot\st{\td}{\mvec n}\!{^{\Box e}}\
\leq\ 0.
\vspace{.3cm}\ee
Without any restriction of generality, the left hand brackets in \R{31} can be presupposed as
being continuous, if the right hand factors vanish. These factors vanish, if suitable
equilibrium environments are chosen for contacting
\bee{32a}
{\cal G}_{\odot}^\Box\ \longrightarrow\ \st{\td}{Q}\!_\odot^\Box = 0, \qquad
{\cal G}_{i0}^\Box\ \longrightarrow\ \st{\td}{n}\!_{i0}^{\Box e} = 0.
\ee
${\cal G}_{\odot}^\Box$ and ${\cal G}_{i0}^\Box$ are equipped with temperatures
$T_\odot^\Box$ and $T_{0}^\Box$ --the same for all $_i^\Box$-components-- with molar enthalpies $h_{i0}^\Box$ and chemical potentials $\mu_{i0}^\Box$. Consequently, according to the proposition \R{30}
\bee{32}
\st{\td}{Q}\!_\odot^\Box\st{*}{=} 0\ \Longleftrightarrow\ \Theta = T^\Box_\odot, 
\qquad
\st{\td}{\mvec n}\!_0^{\Box e} \st{*}{=} \mvec{0}\ \Longleftrightarrow\ 
\Big(\frac{\mvec{h}}{\Theta} = \frac{\mvec{h}^\Box_0}{T_{0}^\Box}\Big)\ \wedge\ 
\Big(\frac{\mvec{\mu}}{\Theta} = \frac{\mvec{\mu}^\Box_0}{T_{0}^\Box}\Big)
\ee
is valid. Here, \R{32}$_2$ holds true for each chemical component. The $^\Box_0$-quantities
are known and belong to the special equilibrium environments \R{32a} which generate the vanishing
RHS factors of \R{31}. 
\vspace{.3cm}\newline
Accepting the inequality \R{29}$_1$ for defining $\Theta$ implies in connection with \R{9a}$_1$
that the conductive entropy flux through $\partial\cal G$ is uncontinuous
\bee{32z}
\frac{\st{\td}{Q}}{\Theta}\ \geq\ -\frac{\st{\td}{Q}\!{^\Box}}{T^\Box}\ \longrightarrow\ 
\frac{\st{\td}{Q}}{\Theta} +\frac{\st{\td}{Q}\!{^\Box}}{T^\Box}\ =:\ \sigma_Q\ \geq\ 0.
\ee
Analogously, the convective fluxes
\bee{32y}
\frac{\mvec{h}}{\Theta}\cdot\st{\td}{\mvec{n}}\!{^e}+\frac{\mvec{h}^\Box}{T^\Box}
\cdot\st{\td}{\mvec{n}}\!{^{\Box e}}\ =:\ \sigma_h\ \geq\ 0,\qquad
-\Big(\frac{\mvec{\mu}}{\Theta}\cdot\st{\td}{\mvec{n}}\!{^e}
+\frac{\mvec{\mu}^\Box}{T^\Box}
\cdot\st{\td}{\mvec{n}}\!{^{\Box e}}\Big)\ =:\ \sigma_\mu\ \geq\ 0
\ee
generate the {\em contact entropy productions} $\sigma_\boxtimes$.

\subsection{Contact temperature}

Using \R{32}$_1$, a non-equilibrium temperature can be introduced by measurement:
Consider a non-equilibrium system $\cal G$ and contact it through an inert partition
$\partial\cal G$ with the special equilibrium environment ${\cal G}^\Box_{\odot}$ which generates
vanishing net heat exchange $\st{\td}{Q}\!_\odot^\Box=0$, then $\cal G$
has by definition \R{32}$_1$ the (non-equilibrium) {\em contact temperature} ${\Theta}$
which is identical with the thermostatic temperature ${T_\odot^\Box}$ of the controlling
${\cal G}^\Box_{\odot}$ \C{MUBR75,MU77,MUBR77}. The concept of contact temperature can also be used, if the contact partition is divided into subsurfaces \C{MUBER04,MUBER07}. Evident is, that the contact temperature depends on the design of the partition $\partial\cal G$. This
influence vanishes in equilibrium, and the contact temperature changes into the thermostatic
temperature.
\vspace{.3cm}\newline
There are a lot of suggestions for non-equilibrium temperatures \C{CAJO03}. But the
problem is to find a thermometer for the proposed non-equilibrium temperature, otherwise it
presents only a quantity of calculation \C{MU14}. The contact temperature is defined by such
a thermometer: this is ${\cal G}_{\odot}^\Box$ having the thermostatic temperature
$T_\odot^\Box =:\Theta$. In more detail:
\begin{center}
\parbox[t]{14cm}{
{\sf Definition:} The system's contact temperature is that thermostatic temperature of the system's equilibrium environment for which the net heat exchange between the system and this environment through an inert partition vanishes by change of sign.} 
\end{center}
As easily to demonstrate, contact temperature $\Theta$ and the internal energy $U$ are independent of each other. For this purpose, a rigid inert partition $\partial\cal G$ ($\st{\td}{\mvec{a}} \equiv\mvec{0}$) is chosen
which is impervious to matter
$(\st{\td}{\mvec{n}}\!{^e}\equiv 0)$ and a time-dependent environment temperature $T^\Box(t)$ which is always set equal to the value of the momentary contact temperature $\Theta(t)$ of $\cal G$: 
\bee{33}
T^\Box (t)\st{*}{=}\Theta (t)\ \longrightarrow\ \st{\td}{Q}\!{^\Box} = -\st{\td}{Q}\ =\ 0\ 
\longrightarrow\ \st{\td}{U}\ =\ 0
\ee
according to \R{17}. Because $\Theta$ is time-dependent and $U$ is constant, totally
different from thermostatics, both quantities are independent of each other.
\vspace{.3cm}\newline
The contact temperature is useful for defining efficiencies which are smaller than the Carnot
efficiency \C{MU08}. Consequently, this non-equilibrium efficiency represents a more realistic
quantity for process evaluation.

\subsection{Non-equilibrium molar enthalpies and chemical potentials}

Using \R{32}$_2$, a non-equilibrium molar entropy can be introduced by measurement:
Consider a non-equilibrium Schottky system $\cal G$ and contact it through an inert partition
$\partial\cal G$ with the special equilibrium environment ${\cal G}^\Box_{ni}$ which generates
vanishing net material exchange of the $_i^\Box$-component $\st{\td}{n_i}\!\!^{\Box e}=0$,
then $\cal G$ has by definition \R{32}$_2$ the {\em non-equilibrium molar enthalpy}
({\em non-equilibrium chemical potential})
\bee{33a}
h_i\ =\ \frac{\Theta}{T_0^\Box}h_{i0}^\Box\ \longrightarrow\ 
\mvec{h}\ =\ \frac{T_\odot^\Box}{T_0^\Box}\mvec{h}_{0}^\Box,
\qquad\Big(\mbox{and analogous }\mvec{\mu}\ =\
\frac{T_\odot^\Box}{T_0^\Box}\mvec{\mu}_{0}^\Box\Big)
\ee
which is identical with the thermostatic molar enthalpy $h_{i0}^\Box$ of the controlling
environment ${\cal G}^\Box_{ni}$, if besides the vanishing material exchange $\st{\td}{n_i}\!\!^{\Box e}=0$
also the heat exchange vanishes
\bee{33b}
T_0^\Box\doteq T_\odot^\Box:\quad
(\st{\td}{\mvec{n}}\!\!^{\Box e}= \mvec{0})\ \wedge\ (\st{\td}{Q}\!_\odot^\Box=0)\ 
\longrightarrow\ \mvec{h}=\mvec{h}_0^\Box,
\qquad\Big(\mbox{and analogous }\mvec{\mu}= \mvec{\mu}_{0}^\Box\Big).
\vspace{.3cm}\ee
Evident is, that different measuring devices (different controlling equilibrium environments)
are necessary for defining the contact temperature $\Theta$, the non-equilibrium
molar enthalpies $h_i$ and chemical potentials $\mu_i$ of the chemical components in $\cal G$.

\subsection{Non-equilibrium molar entropies}

According to \R{31z}$_2$, \R{33a} and \R{33b}$_1$, the non-equilibrium molar entropy is
\bee{33d}
\mvec{s}\ =\ \frac{1}{\Theta}\Big(\mvec{h}-\mvec{\mu}\Big)\ =\
\frac{1}{T_0^\Box}\Big(\mvec{h}_0^\Box-\mvec{\mu}_0^\Box\Big)\ =\ \mvec{s}_0^\Box.
\ee
The non-equilibrium molar entropy is defined by the thermostatic molar entropy of that equilibrium
environment that generates vanishing material exchange.
\vspace{.3cm}\newline
Here is the synopsis of the contact quantities, their defining quantities and the corresponding
controlling equilibrium environments:
\byy{34}
\cal G&\quad\longrightarrow\quad& \Theta,\mvec{h},\mvec{\mu},\mvec{s}\qquad\mbox{non-equilibrium}
\\ \label{35}
\cal G^\Box&\quad\longrightarrow\quad& T^\Box,\mvec{h}^\Box,\mvec{\mu}^\Box,\mvec{s}^\Box\qquad^\Box\mbox{equilibrium environment}
\\ \label{36}
\cal G^\Box_\odot&\quad\longrightarrow\quad& T_\odot^\Box\qquad\mbox{vanishing heat exchange}
\\ \label{37}
{\cal G}^{\Box}_{ni}&\quad\longrightarrow\quad&T_0^\Box,h_{i0}^\Box,\mu_{i0}^\Box, s_{i0}^\Box
\qquad\mbox{vanishing material exchange}
\eey
The quantities $\boxtimes_{\odot,0}^\Box$ marked by $\odot$ or zero are given thermostatic
quantities of several equilibrium environments. The resulting contact quantities of \R{34} are
determined by the following equations: \R{32}$_{1,2}$ and \R{33d}. The construction of a
non-equilibrium entropy rate can now go on.

\section{Verification: Non-Equilibrium Entropy}

Starting with the preliminary shape of the entropy rate \R{21}, the contact temperature $\Theta$
is now defined by \R{32}$_1$ and the non-equilibrium molar entropy by \R{33d}. The shape of
the entropy production $\Sigma$ is still missing. Beyond that, the entropy rate \R{21} is up to
now no time derivative of a state function of the Schottky system $\cal G$ (oldfashioned: no total differential of something), but only a time rate along a process. Consequently, a suitable
non-equilibrium state space has to be specified in \R{11}$_2$ which allows to generate a
non-equilibrium entropy as a state function on it.

\subsection{A non-equilibrium state space}

Because the equilibrium subspace \R{11}$_1$ is spanned by the work variables $\mvec{a}$, the
mole numbers $\mvec{n}$ and the internal energy $U$, these state variables appear also in the non-equilibrium
state space. If the First Law \R{17} and \R{31z}$_2$ are inserted, \R{21} results in
\bee{38}
\st{\td}{S}\ =\ \frac{1}{\Theta}\Big(\st{\td}{U}-\mvec{h}\cdot\st{\td}{\mvec{n}}\!{^e}
-{\bf A}\cdot\st{\td}{\mvec{a}}+\mvec{h}\cdot\st{\td}{\mvec{n}}\!{^e}-
\mvec{\mvec{\mu}}\cdot\st{\td}{\mvec{n}}\!{^e}\Big)+\Sigma.
\ee
Because the external mole number rates $\st{\td}{\mvec{n}}\!{^e}$ are no state variables, 
but the mole numbers themselves are included in the equilibrium subspace \R{11}$_1$
according to the Zeroth Law, the missing term for generating the mole numbers in \R{38} is hidden in the entropy production
\bee{39}
\Sigma\ =\ -\frac{1}{\Theta}\mvec{\mvec{\mu}}\cdot\st{\td}{\mvec{n}}\!{^i}
+\Sigma^0,
\ee
and taking \R{18} into account, \R{38} results in
\bee{40}
\st{\td}{S}\ =\ \frac{1}{\Theta}\Big(\st{\td}{U}
-{\bf A}\cdot\st{\td}{\mvec{a}}-
\mvec{\mvec{\mu}}\cdot\st{\td}{\mvec{n}}\Big)+\Sigma^0.
\ee
Because the bracket in \R{40} contains only equilibrium variables,
the non-equilibrium state variables appear in the entropy production $\Sigma^0$. 
Because the contact temperature is independent of the internal energy, it represents an
additional variable which is included in $\Sigma^0$. The choice of further non-equilibrium variables
depends on the system in consideration. Here, {\em internal variables} $\mvec{\xi}$ are chosen because they allow a great flexibility in describing non-equilibria \C{MU90,MUMAU94}.
Consequently, the created non-equilibrium state space is
\bee{41}
Z\ =\ (\mvec{a},\mvec{n},U,\Theta,\mvec{\xi})
\ee
(this is an example, other state spaces are of course possible).
The entropy rate \R{40} and the entropy production with regard to the non-equilibrium variables
$\Theta$ and $\mvec{\xi}$ become
\bee{42}
\st{\td}{S}(Z)\ =\ \frac{1}{\Theta}\Big(\st{\td}{U}
-{\bf A}\cdot\st{\td}{\mvec{a}}-
\mvec{\mvec{\mu}}\cdot\st{\td}{\mvec{n}}\Big)
+\alpha\st{\td}{\Theta}+\mvec{\beta}\cdot\st{\td}{\mvec{\xi}}\ \longrightarrow\ \Sigma^0 =
\alpha\st{\td}{\Theta}+\mvec{\beta}\cdot\st{\td}{\mvec{\xi}}\ \geq\ 0.
\vspace{.3cm}\ee
The time rate along a process on a state space does not necessarily belong to a state function:
two additional requirements must be satisfied, firstly the {\em embedding theorem} which guarantees that the so constructed non-equilibrium entropy time rate is in accordance with the
presupposed equilibrium entropy, and secondly the {\em adiabatical uniqueness} which enforces
that the time rate of the non-equilibrium entropy is a total differential of the state space variables.

\subsection{Embedding theorem}

The time rate of the non-equilibrium entropy has to be in accordance with the --as known presupposed-- equilibrium entropy: the  non-equilibrium entropy rate integrated along an irreversible process $\cal T$ starting and ending in equilibrium states --$A_{eq}$ and $B_{eq}$--
has the same value as the equilibrium entropy difference between these two equilibrium states
calculated along the corresponding accompanying process $\cal R$ \R{21w}
\bee{43}
{\cal T}\int_{Aeq}^{Beq}\st{\td}{S}(Z)dt\ =\ 
{\cal R}\int_{Aeq}^{Beq}\st{\td}{S}\Big({\cal P}(Z)\Big)dt\ =\ S(B_{eq})-S(A_{eq}).
\ee
Taking the projection \R{21w} and the entropy production \R{42}$_2$ into account, \R{43} results in
\bey\nonumber
0\ &=& {\cal T}\int_{Aeq}^{Beq}\Big(\st{\td}{S}(Z)-\st{\td}{S}({\cal P}(Z))\Big)dt\ \geq\
\\ \label{44}
&\geq& {\cal T}\int_{Aeq}^{Beq}\Big[\Big(\frac{1}{\Theta}-\frac{1}{T^*}\Big)\st{\td}{U}-
\Big(\frac{\bf A}{\Theta}-\frac{\bf A^*}{T^*}\Big)\cdot\st{\td}{\mvec{a}}-
\Big(\frac{\mvec{\mu}}{\Theta}-\frac{\mvec{\mu}^*}{T^*}\Big)\cdot\st{\td}{\mvec{n}}\Big]dt.
\eey
This inequality reminds of the "Fundamentale Ungleichung" \C{MEI69} of the "Entropiefreie Thermodynamik" \C{MEI69a,KEL72} because an entropy does not appear in \R{44}.
Of course, sixty years later the formal background is different: first of all, the realization
that the introduction of a non-equilibrium entropy also requires an introduction of a well defined
non-equilibrium temperature, that reversible accompanying processes arise from projections on
the equilibrium subspace and that a non-equilibrium large state space is necessary for defining
a non-equilibrium entropy as a state function.
\vspace{.3cm}\newline
Consider a cyclic process which at least contains one equilibrium state $A_{eq}$, \R{43} results in
\bee{45}
B_{eq}\succapprox A_{eq}:\qquad ^{(A_{eq})}\!\!\oint\st{\td}{S}(Z)dt\ =\ 0,
\ee
a relation which does not enable the entropy rate to be a total differential on the state space
because this special cyclic processes contains at least one equilibrium state. A generalization is
made in the next section.

\subsection{Adiabatical uniqueness}

Consider an arbitrary, but fixed non-equilibrium state $B$ and a process family whose processes $\cal T$ all start at different arbitrary equilibrium states $A_{eq}$ and end in $B$. Subsequently,
an adiabatic process takes place starting from $B$ and ending in an equilibrium state $C_{eq}$
\byy{46}  
{\cal T}:\ A_{eq}&\longrightarrow& B,\quad\mbox{different $A_{eq}$, arbitrary ${\cal T}$, fixed $B$},
\\ \label{47}
{\cal A}:\ B&\longrightarrow& C_{eq},\quad\mbox{fixed $B$, adiabatic ${\cal A}$}.  
\eey
The entropy change along these processes is according to the embedding theorem
\byy{48}
&&{\cal T}\int_{A_{eq}}^B \st{\td}{S} dt + {\cal A}\int_B ^{C_{eq}}\st{\td}{S}dt \ =\
S_{eq}^C  - S_{eq}^A,
\\ \label{49}
&& {\cal T}\int_{A_{eq}}^B \st{\td}{S} dt\ =\ S^B(A_{eq},{\cal T})-S_{eq}^A.
\eey
The non-equilibrium entropy $S^B$ may depend on ${\cal T}$ and its starting state $A_{eq}$.
Inserting \R{49} into \R{48} results in
\bee{50}
S^B(A_{eq},{\cal T})\ =\ S_{eq}^C-{\cal A}\int_B ^{C_{eq}}\st{\td}{S}dt.
\ee
If the final equilibrium state $C_{eq}$ in which the adiabatic process ends does not depend on
different $(A_{eq},{\cal T})$, also the entropy $S^B$ of the fixed $B$ does not depend on
them and the LHS of \R{50} represents a process independent non-equilibrium entropy
(not only a rate) of arbitrary chosen non-equilibrium states $B$ whose value is given by the
RHS of \R{50}, if $C_{eq}$ is unique. 
\vspace{.3cm}\newline
This uniqueness is satisfied in phenomenological, but not in stochastic thermodynamics \C{MU16}.
In more detail, the condition runs as follows \C{MU09}
\begin{center}
\parbox[t]{14cm}{
{\sf Definition:} A Schottky system is called {\em adiabatically unique}, if for each arbitrary, but fixed
non-equilibrium state $B$ after isolation of the system, the relaxation process ends always in the same final equilibrium state, independently of how the process into $B$ was performed.}
\end{center}
Consequently, a non-equilibrium entropy of Schottky systems 
\bee{51}
S^B(Z)\ =\ S_{eq}^C-{\cal A}\int_B ^{C_{eq}}\st{\td}{S}dt
\ee
can always be defined as a state function on a non-equilibrium state space, if the system is adiabatically unique. This well founded definition is evidently more than a "primitive concept"
to which Meixner objected. Also along a non-equilibrium process, the entropy time rate inequality
\bee{52}
\st{\td}{S}\ \geq\ \frac{1}{\Theta}\st{\td}{Q}+\mvec{s}\cdot\st{\td}{\mvec n}\!{^e}
\ee
is in this case valid according to \R{21} and \R{28}. Consequently, one of the conditions which
Meixner missed when he stated \C{MEI67} "there is no entropy along an irreversible process" is valid and results in a generalization of Clausius inequality
\bee{53} 
\oint\st{\td}{S}dt\ =\ 0\ \geq\ \oint\Big(\frac{1}{\Theta}\st{\td}{Q}
+\mvec{s}\cdot\st{\td}{\mvec n}\!{^e}\Big)dt
\ee
in which no quantities of the controlling environment appears. The original Clausius inequality for open systems results from use of the defining inequalities \R{29}
\byy{54} 
 0\ \geq\ \oint\Big(\frac{1}{\Theta}\st{\td}{Q}
+\mvec{s}\cdot\st{\td}{\mvec n}\!{^e}\Big)dt\ \geq\ 
\oint\Big(\frac{1}{T^\Box}\st{\td}{Q}+\mvec{s}^\Box\cdot\st{\td}{\mvec n}\!{^e}\Big)dt\
=\ SL,
\\ \label{54a}
\oint\Big[\Big(\frac{1}{\Theta}-\frac{1}{T^\Box}\Big)\st{\td}{Q}
+(\mvec{s}-\mvec{s}^\Box )\cdot\st{\td}{\mvec n}\!{^e}\Big]dt\ \geq\ 0.
\hspace{2.5cm}
\eey
The cyclic path integral runs on the non-equilibrium state space. Consequently, the four questions
posed at the beginning of the paper are answered. Especially, the reasoning
--defining inequalities, non-equilibrium entropy, extended and ordinary Clausius inequality--
elucidates how to answer the questions. Reversible or better accompanying processes are by
projection generated mathematical pathes on the equilibrium subspace. Unusual is, that internal
energy and contact temperature are independent state variables, the first an equilibrium
variable and the second a non-equilibrium one. Conclutions of this fact are discussed in the next section.

\subsection{The integrability conditions}

If a Schottky system ${\cal G}$ is adiabatically unique, a non-equilibrium entropy exists according
to \R{51}
\bee{55}
S\ =\ S(U,\mvec{a},\mvec{n}, \Theta,\mvec{\xi}),
\ee
and the path integrals on the non-equilibrium state space over the entropy rate between two
states are path independent. Because \R{55} is a state function, from \R{42}$_1$ follows
for the partial derivatives of the entropy
\byy{56}
\frac{\partial S}{\partial U}\ =\ \frac{1}{\Theta},\quad
\frac{\partial S}{\partial \mvec{a}}\ =\ -\frac{\mvec{A}}{\Theta},\quad
\frac{\partial S}{\partial \mvec{n}}\ =\ -\frac{\mvec{\mu}}{\Theta},
\\ \label{57}
\frac{\partial S}{\partial \Theta}\ =\ \alpha ,\quad
\frac{\partial S}{\partial \mvec{\xi}}\ =\ \mvec{\beta}.  
\eey
Because $\Theta$ and $U$ are independent of each other according to \R{33}, we can integrate
\R{56}$_1$ immediately
\bee{58}
S(U,\mvec{a},\mvec{n},\Theta , \mvec{\xi})\ =\ 
\frac{1}{\Theta}U + K(\mvec{a},\mvec{n},\Theta , \mvec{\xi}).
\ee
Consequently, the non-equilibrium entropy is a linear function of the internal energy. Here
\bee{59}
U-\Theta S\ =\ 
-\Theta K\ =:\ F(\mvec{a},\mvec{n},\Theta , \mvec{\xi})
\ee
is the free energy $F$ of ${\cal G}$. 
\vspace{.3cm}\newline
Because of \R{10a} and \R{19w}, in equilibrium is valid
\bee{60}
S^{*}\ =\ \frac{1}{T^*}U - \frac{F}{T^*}\Big(\mvec{a},\mvec{n} ,\Theta(U, \mvec{a}, \mvec{n}),\mvec{\xi}(U,\mvec{a}, \mvec{n})\Big),
\ee
an expression which is in general non-linear in $U$. Consequently, the equilibrium entropy
$S^*$ is in contrast to the non-equilibrium entropy \R{58} non-linear in $U$.
From the integrability conditions \R{56} and \R{57} follows that except of
$\alpha$ all constitutive equations do not depend on the internal energy $U$ in non-equilibrium:
\byy{61}
\frac{\partial}{\partial\mvec{a}}\frac{\partial S}{\partial U}&=& \mvec{0}
\quad\Longrightarrow\quad \frac{\partial\mvec{A}}{\partial U}\ =\ \mvec{0},
\\ \label{62}
\frac{\partial}{\partial\mvec{n}}\frac{\partial S}{\partial U} &=& \mvec{0}
\quad\Longrightarrow\quad \frac{\partial\mvec{\mu}}{\partial U}\ =\ \mvec{0},
\\ \label{63}
\frac{\partial}{\partial\mvec{\xi}}\frac{\partial S}{\partial U}&=& \mvec{0}
\quad\Longrightarrow\quad \frac{\partial\mvec{\beta}}{\partial U}\ =\ \mvec{0},
\\ \label{64}
\frac{\partial}{\partial\Theta}\frac{\partial S}{\partial U}&=&
-\frac{1}{\Theta ^2}\ =\  
\frac{\partial\alpha}{\partial U}\ \longrightarrow\ \alpha\ =\ -\frac{U}{\Theta ^2}.
\eey
\vspace{.3cm}\newline
Apart from these specialities caused by the independence of the contact temperature from the
internal energy, the formal structure of phenomenological non-equilibrium thermodynamics is
except of the entropy production very similar to thermostatics.

\section{Summary}

Meixner's historical remark \C{MEI67} {\em "... it can be shown that the concept of entropy in the absence of equilibrium is in fact not only questionable but that it cannot even be defined"}
was the starting-point of a deliberation, if there is a possibility to introduce non-equilibrium entropies
which are better founded than a primitive concept. Here, steps are done for a basic
substantiation of non-equilibrium entropies for Schottky systems (concerning field formulation see
\C{MU04a,WAHU18}). The chain of reasoning is as follows:
\begin{itemize}   
\item Introduce a large state space \C{MUAS1} which is composed of an equilibrium subspace
and a non-equilibrium part 
\item The variables of the equilibrium subspace are determined by the Zeroth Law: work variables,
mole numbers and internal energy
\item Processes are trajectories on the state space
\item Reversible processes are projections of non-equilibrium processes onto the equi\-li\-brium
subspace. An accompanying process \C{KE71} with time as a path parameter is generated by
projection of the corresponding non-equilibrium process onto the equilibrium subspace 
\item The time rate of the internal energy is introduced by the First Law
\item The entropy is in equilibrium and in non-equilibrium a balanceable quantity
\item The entropies of partial systems are additive
\item Introduction of the Second Law for isolated systems and entropy productions \C{MU04}
\item The defining inequalities for contact temperature, non-equilibrium molar enthalpies and
chemical potentials resulting in the non-equilibrium molar entropy
\item The embedding theorem enforcing compatibility of a non-equilibrium entropy with the
equilibrium one
\item Adiabatic uniqueness guaranteeing that the non-equilibrium entropy is a state function on
the non-equilibrium state space.
\end{itemize}
These items make possible to generate a non-equilibrium entropy as a state state function.
Beyond that, the Clausius inequality for open systems \R{54}$_2$ follows including thermostatic
temperature and equilibrium molar entropy of the system's controlling environment. Additionally,
the four questions of the introduction are answered.
\vspace{.3cm}\newline
Fifty years ago, the thermodynamical society was not aware that the above mentioned items
are necessary to get rid of the primitive concept of a non-equilibrium entropy. Thus, it is evident
that a more axiomatically oriented thermodynamicist was endeavoured at that time to avoid the use of a non-equilibrium entropy.

\section{Closure}

Non-equilibrium open Schottky systems are characterized by contact quantities whose definitions
require inert partitions. If contact quantities seem to be too artificial, the non-equilibrium Schottky
system can be approximatively replaced by an equilibrium one which is described by the
accompaying process \R{21w} of the Schottky system resulting in the contact of two equilibrium
systems. This kind of description is called {\em endoreversible thermodynamics}. The
non-equilibrium variable contact temperature becomes dependent on the internal energy and the
other equilibrium variables, known as caloric equation of state. The entropy productions in the
Schottky system and its environment vanish, but the contact entropy productions remain
because the contact problem is unchanged. The non-equilibrium contact quantities of the
Schottky system are replaced by the equilibrium quantities of the accompanying process
resulting in the contact entropy productions \R{32z} and \R{32y}. Endoreversible
thermodynamics is analogous to the {\em hypothesis of local equilibrium} in field theories of
thermodynamics. 
\vspace{1cm}\newline
{\bf Acknowledgement:} Vivid discussions with J.U. Keller and Christina Papenfuss are gratefully acknowledged.


\begin{thebibliography}{99}


\bibitem{SCHO29} W. Schottky: Thermodynamik, Erster Teil $\S 1$, Springer, Berlin 1929

\bibitem{Kelvin}
Mathematical and Physical Papers of William Thomson. Cambridge University Press, Cambridge
1882, Vol.1: 100-106, 113-140, 174-200

\bibitem{Clausius}
R. Clausius: On a modified form of the second Fundamental Theorem in the Mechanical
Theory of Heat, Phil. Mag., ser. 4, 12 (1854), 81; translated from Pogg. Ann., 93 (1854), 481.

\bibitem{MUAS}
W. Muschik: Aspects of Non-Equilibrium Thermodynamics. World Scientific, Singapore 1990; 3.4.1

\bibitem{Carnot}
S. Carnot: R\'{e}flexions sur la puissance motrice du feu sur les machines. Bachelier, Paris 1824

\bibitem{MU89}
W. Muschik: Thermodynamical algebra, Second Law, and Clausius' inequality at negative absolute
temperatures. Journal of Non-Equilibrium Thermodynamics 14  (1989) 173-198

\bibitem{Hutter}
 K. Hutter: The foundations of thermodynamics, its basic 
postulates and implications. A review of modern thermodynamics,
Acta Mechanica 27 (1977) 1-54

\bibitem{KE76} J. Kestin (Ed): The Second Law of Thermodynamics,
  Dowden, Hutchinson and Ross, Stroudsburg, 1976 

\bibitem{SER79}
J. Serrin: Conceptual Analysis of the Classical Second Laws of Thermodynamics.
Arch. Rat. Mech. Anal. 70 (1979) 355-371

\bibitem{SIL83}
M. Silhavy: On the Clausius Inequality. Arch. Rat. Mech. Anal. 81 (1983) 221-243

\bibitem{MU88}
W. Muschik: Formulations of the Second Law - Recent Developments.
Journal of Physics and Chemistry of Solids 49 (1988) 709-720

\bibitem{MU90b} W. Muschik: Second law: Sears-Kestin statement and
Clausius inequality. Am. J. Phys. 58 (1990) 241-244

\bibitem{MUEH96} W. Muschik, H. Ehrentraut: An amendment to the second
 law. J. Non-Equilib. Thermodyn. 21 (1996) 175-192

\bibitem{MU04}
W. Muschik: Different Formulations of the Second Law. In: B.T. Maruszewski, W. Muschik, A. Radowicz
(Eds), Proceedings of the International Symposium on Trends in Continuum Physics (Trecop '04),
p.1 - 12, Publishing House of Poznan University of Technology, 2004, ISBN 83-7143-297-6

\bibitem{MU09} W. Muschik: Contact quantities and non-equilibriun entropy of
discrete systems.  J. Non-Equilib. Thermodyn. 34 (2009) 75-92

\bibitem{MUBER04} W. Muschik, A. Berezovski: 
Thermodynamic interaction between two discrete systems in non-equilibrium.
J. Non-Equilib. Thermodyn. 29 (2004) 237-255

\bibitem{MUBER07} W. Muschik, A. Berezovski: Non-equilibrium contact quantities and
compound deﬁciency at interfaces between discrete systems. 
Proc. Estonian Acad. Sci. Phys. Math. 56 (2007) 133–145

\bibitem{MUAS1}
W. Muschik: Aspects of Non-Equilibrium Thermodynamics. World Scientific, Singapore 1990; 1.2

\bibitem{KE71} J.U. Keller: Ein Beitrag zur Thermodynamik fluider Systeme.
Physica 53 (1971) 602-620

\bibitem{MUGU99} W. Muschik, S. G\"umbel: 
Does Clausius' inequality analogue exists for open discrete systems ?.
J. Non-Equilib. Thermodyn. 24 (1999) 97-106

\bibitem{HA69} R. Haase: Thermodynamics of Irreversible Processes, §1.7 Addison-Wesley, Reading Ma. 1969

\bibitem{BO21} M. Born: Kritische Betrachtungen zur Darstellung der Thermodynamik.
Physikalische Zeitschrift 22 (1921) 218-224; 249-254; 282-286

\bibitem{MEI67} J. Meixner: Beziehungen zwischen Netzwerktheorie und Thermodynamik.
Arbeitsgemeinschaft f\"ur Forschung des Landes Nordrhein-Westfalen, Heft 181, Westdeutscher
Verlag, K\"oln/Opladen 1968; Supplement

\bibitem{MEI68} J. Meixner: Entropie im Nichtgleichgewicht. Rheologica Acta 7 (1968) 8-13

\bibitem{MEI69} J. Meixner: Thermodynamik der Vorg\"ange in einfachen fluiden Medien und
die Charakterisierung der Thermodynamik irreversibler Prozesse. Z. Physik 219 (1969) 79-104

\bibitem{KEL72} J.U. Keller: \"Uber den 2. Hauptsatz der Thermodynamik irreversibler Prozesse.
Acta Physica Austriaca 35 (1972) 321-330

\bibitem{KER72} W. Kern: Zur Vieldeutigkeit der Gleichgewichtsentropie in kontinuierlichen Medien.
Dissertation, RWTH Aachen 1972

\bibitem{KES79} J. Kestin: A Course in Thermodynamics, Vol.I. Hemisphere Pub. Corp.,
Washington/London 1979; sect.13.6

\bibitem{MU84} W. Muschik: Recent developments in nonequilibrium thermodynamics, in: Lecture
Notes in Physics, Vol.199, p. 387, Springer, Berlin 1984 

\bibitem{MUBR75} W. Muschik, G. Brunk: Temperatur und Irreversibilit\"at in der Rationalen
Mechanik. ZAMM 55 (1975) T102-T105

\bibitem{MU77} W. Muschik: Empirical foundation and axiomatic treatment of
non-equilibrium temperature. Arch. Rat. Mech.Anal. 66 (1977) 379-401

\bibitem{MUBR77} W. Muschik, G. Brunk: A concept of non-equilibrum temperature.
Int. J. Eng. Sci. 15 (1977) 377-389 

\bibitem{CAJO03} J.Casas-V\'azquez, D.Jou: Temperature in non-equilibrium states: 
A review of open problems and current proposals. Rep.Prog.Phys. 66 (2003) 1937-2023

\bibitem{MU14} W. Muschik: Contact temperature and internal variables:
A glance back, 20 years later.
J. Non-Equilib. Thermodyn. 39 (2014) 113-121

\bibitem{MU08} W. Muschik: Survey of some branches of thermodynamics.
J. Non-Equilib. Thermodyn. 33 (2008) 165–198; 2.1.2

\bibitem{MU90} W. Muschik: Internal variables in non-equilibrium thermodynamics.
J. Non-Equilib. Thermodyn. 15 (1990) 127-137

\bibitem{MUMAU94} G.A. Maugin, W. Muschik: Thermodynamics with internal variables.
J. Non-Equilib. Thermodyn. 19 (1994) 217 - 249; 250-289

\bibitem{MEI69a} J. Meixner: Processes in simple thermodynamic materials.
Arch. Rat. Mech. Anal. 33 (1969) 33-53

\bibitem{MU16} W. Muschik: Non-equilibrium thermodynamics and stochasticity: a phenomenological look on Jarzynski's equality.
Continuum Mechanics and Thermodynamics 28 (2016) 1887-1903

\bibitem{MU04a} W. Muschik: Remarks on thermodynamical terminology.
Journal of Non-Equilibrium Thermodynamics 29 (2004) 199-203

\bibitem{WAHU18} Y. Wang, K. Hutter: Phenomenological thermodynamics of  irreversible
processes. Entropy 2018, 20, 479 
 

\end{thebibliography}
\end{document}